# A RELATIVISTIC-PROTON DARK MATTER WOULD BE EVIDENCE THE BIG BANG PROBABLY SATISFIED THE SECOND LAW OF THERMODYNAMICS

Jerome Drexler
Former NJIT Research Professor of Physics
Member, Board of Overseers
New Jersey Institute of Technology

**Abstract**

A new research hypothesis has been developed by the author based upon finding astronomically based "cosmic constituents" of the Universe that may be created or influenced by or have a special relationship with possible dark matter (DM) candidates. He then developed a list of 14 relevant and plausible "cosmic constituents" of the Universe, which then was used to establish a list of constraints regarding the nature and characteristics of the long-sought dark matter particles. A dark matter candidate was then found that best conformed to the 14 constraints established by the "cosmic constituents." The author then used this same dark matter candidate to provide evidence that the Big Bang was relativistic, had a low entropy, and therefore probably satisfied the Second Law of Thermodynamics.

**Determining The Nature Of The Dark Matter Of The Universe**

One hundred years ago, Albert Einstein announced the Special Theory of Relativity, which predicted and explained that a proton traveling near the speed of light could have a relativistic mass a thousand, a million, or even a billion times greater than the mass of a proton at rest. Therefore, the gravitational strength of the multitudinous galaxy-orbiting relativistic protons moving in the cosmos could create extremely large gravity-related tidal forces on nearby matter, like that exhibited by dark matter.

Astronomer Fritz Zwicky [1] discovered the presence of dark matter in the Coma cluster of galaxies in 1933. Ever since astronomer Vera Rubin [2,3] confirmed the existence of dark matter halos around galaxies in 1977, cosmologists and astrophysicists have been trying to identify the dark matter particles.

In 1984, scientists [4] developed a Cold Dark Matter (CDM) theory based upon a theoretical uncharged, slow moving particle that they called the Weakly Interacting Massive Particle (WIMP). More recently, it was estimated by scientists that the theoretical WIMP dark matter particles would require a mass in the range of about 35 to 10,000 times [5] greater than the mass of a proton at rest in order to exhibit the observed gravity-related forces of dark matter halos. However, searches for the theoretical WIMP particles during the past 20 years have all come up empty handed.

For this reason, and knowing that Einstein's relativistic proton easily could meet the mass requirement of the mysterious dark matter particles and that relativistic cosmic ray protons are widely observed, the author has endeavoured to determine the nature of dark matter.

The author posits that the galaxy-orbiting relativistic proton has the necessary characteristics of the long-sought dark matter particles, which are estimated by most scientists to comprise 80% to 90% [6] of the total mass of the Universe. Relativistic protons do have the required mass and the required difficulty of detection. Protons also can transform themselves into hydrogen, the principal matter of galaxies, by creating muons [7,8] that decay into electrons, then combining with the electrons.

Thus, relativistic protons could form (1) galaxies and their dark matter halos, (2) galaxy clusters and their dark matter halos, and (3) the long, large, filamentary dark matter known to crisscross the cosmos.

However, for this proton-based dark matter theory to become widely accepted, there also should be astronomical evidence of relativistic protons within the dark matter halo surrounding the Milky Way. The author posits that the high-energy cosmic ray relativistic protons bombarding Earth every day, uniformly from all directions, lend credence toward providing such astronomical evidence.

The author has applied a cryptographic-like analysis for solving the mystery of the identity of dark matter of the Universe. Instead of using an encrypted message to extract the secret code it contains as in normal cryptography, the author used 14 cosmic constituents of the Universe to extract the nature and identity of dark matter.

The author had speculated that if dark matter represents 80% to 90% of the mass of the Universe, dark matter should have roles, functions or an influence on most of the following 14 cosmic constituents. Each type of dark matter proposed by scientists was subjected to 14 elimination tests as follows.

The author asked 14 rhetorical questions: Which type of dark matter (DM) particles could:

1. Form spherical dark matter halos around galaxies and DM halos around galaxy clusters?

2. Cause the accelerating expansion of the Universe and possibly store dark energy?

3. Be transformed into low-velocity hydrogen, protons, or proton cosmic rays?

4. Create the magnetic fields within and around spiral galaxies?

5. Be concentrated in the long, large, curved filaments of dark matter, announced by NASA on September 8, 2004, which form galaxy clusters where two DM filaments intersect?

6. Create large, mature, spiral galaxies less than 2.5 billion years after the Big Bang?

7. Create spherical DM halos having predictable outer and "hollow" core diameters?

8. Provide angular momentum to spiral galaxies and DM halos?

9. Create galaxies without a central DM density cusp?

10. Create a starless galaxy or a Low Surface Brightness (LSB) dwarf galaxy with low star formation rates?

11. Lead to linearly rising rotation curves for LSB dwarf galaxies and to flat rotation curves for spiral galaxies?

12. Form 80% to 90% of the mass of the Universe, the remainder being hydrogen, helium, etc.?

13. Ignite hydrogen fusion reactions of second generation stars utilizing hydrogen molecules and dust and ignite fusion reactions of the first generation stars with only hydrogen atoms?

14. Create the first "knee" at $3 \times 10^{15}$ eV, the second "knee" between $10^{17}$ eV and $10^{18}$ eV, and the "ankle" at $3 \times 10^{18}$ eV of the cosmic-ray energy spectrum near the Earth?

After careful study and analysis, the author concluded that galaxy-orbiting relativistic protons would provide many more affirmative answers to the 14 questions than any other known particle. Therefore, relativistic-proton dark matter could be the identity of dark matter since it appears to have the strongest influence on and relationship with the 14 cosmic constituents. The above dark matter identification procedure could also be described as utilizing Ockham's (Occam's) Razor logic 14 times.

Relativistic-proton dark matter satisfies the three basic requirements of a dark matter candidate. Do such protons have sufficient mass? Yes, relativistic protons can have enormous mass. Have they ever been detected? Yes, relativistic protons bombard the Earth every day and are called cosmic rays. Don't relativistic protons move too fast to form small galaxies? The protons can form small galaxies after the protons are slowed down by muon-producing *[7,8]* collisions and synchrotron emission energy losses, and after the protons combine with the electrons created by the muon decay, thereby forming hydrogen.

Since protons are electrically charged particles, they would be constrained by the weak extragalactic and galactic magnetic fields into extremely large circular/spiral orbits forming dark matter halos around galaxies and around galaxy clusters and also could be concentrated into long large curved filaments of dark matter. All three of these dark matter configurations have been reported by astronomers.

Much of the above information was derived from the author's May 2006 book *[8]* and his 19-page April 2005 paper, "Identifying Dark Matter Through the Constraints Imposed by Fourteen Astronomically Based Cosmic Constituents" *[9]*, found on the arXiv.gov website as e-print No. astro-ph/0504512.

The author's 295-page May 2006 book, "Comprehending and Decoding the Cosmos," analyzes an additional 11 cosmic enigmas beyond the 14 derived from his astro-ph paper. Utilizing only relativistic-proton dark matter and the laws of physics, the author explains in a plausible manner all 11 of these recently discovered cosmic enigmas, further supporting the relativistic-proton dark matter theory.

The author's research has led not only to the identification of the dark matter but also to the discovery of the surprising and significant roles and functions of dark matter in creating spiral galaxies, stars, starburst galaxies, extreme ultraviolet synchrotron radiation, and the ultra-high-energy cosmic rays that bombard the Earth. *[8]* Dark matter appears to be a very active and dynamic medium comprising relativistic protons and helium nuclei in the well-known ratio of 12 to 1. Dark matter is widely believed to represent 80% to 90% of the mass of the Universe and believed to be created by the Big Bang. These dark matter characteristics provide the evidence required and used in the next section to reach the conclusion that the Big Bang was relativistic, had a low entropy, and probably satisfied the Second Law of Thermodynamics.

**A Relativistic-Proton Dark Matter Would Be Evidence That The Big Bang Had Low Entropy And Probably Satisfied The Second Law Of Thermodynamics**

In a surprising manner, the Big Bang may have satisfied the Second Law of Thermodynamics. An understanding of this phenomenon is helped by an excerpt from Stephen Hawking's earlier tutorial *[10]* on the subjects of disorder, entropy, the Second Law of Thermodynamics, and the arrow of time: "It is a matter of common experience, that things get more disordered and chaotic with time. This observation can be elevated to the status of a law, the so-called Second Law of Thermodynamics. This says that the total amount of disorder, or entropy, in the universe, always increases with time."

If the amount of disorder, or entropy, in the Universe always increases with time, then at the beginning of time the entropy must have been at its lowest level. The Big Bang also occurred at the beginning of time. Therefore, if we accept the Second Law of Thermodynamics, we must also accept that shortly after the Big Bang the entropy of the Universe would be at the lowest level it would reach throughout all time.

However, the Big Bang is normally characterized as a fiery, chaotic, and massive explosion associated with a high level of disorder and entropy. We are thus faced with an enigma/dilemma as to the level of entropy following the Big Bang, but we are not alone.

On November 18, 2004, the University of Chicago published an article *[11]* entitled, "Astrophysicists attempt to answer the mystery of entropy," that contains the following relevant two-sentence paragraph: "But the mystery remains as to why entropy was low in the universe to begin with. The difficulty of that question has long bothered scientists, who most often simply leave it as a puzzle to answer in the future."

If the entropy following the Big Bang had been very low, the Second Law of Thermodynamics would have been satisfied, but how could a fiery, chaotic Big Bang explosion have a low entropy? This is the enigma that "has long bothered scientists."

The author sees a possible solution to this enigma that would have the Big Bang firing out, in all directions, high-speed ultra-high-energy (UHE) relativistic protons and helium nuclei in the well-known ratio of 12 to 1. A very high percentage of their energies would be available to do work in the Universe while their entropy, the measure of the amount of their energy which is unavailable to do work, would be very low. Such a Big Bang, characterized by a dispersion of UHE relativistic nuclei, could create very high usable energy and very low entropy, and could be designated a Relativistic Big Bang (RBB).

The temperature of a Relativistic Big Bang could be estimated by averaging the energies of the relativistic protons and helium nuclei. The estimated temperature probably would be of the same order of magnitude as the temperature scientists estimate for the Big Bang. Nevertheless, the Relativistic Big Bang would have the very low entropy that the Second Law of Thermodynamics requires for the "beginning of time."

Some astronomical evidence for a Relativistic Big Bang comes from the ultra-high-energy cosmic ray (UHECR) protons that bombard the Earth every day. The RBB is the most plausible origin of these UHECRs. In the author's relativistic-proton DM theory, these UHECRs are stragglers from the galaxy-orbiting UHE relativistic protons that form the dark matter streams in the halos surrounding galaxies.

It is widely accepted that the mass of dark matter today totals about 83% of the mass of the Universe and that dark matter was created by the Big Bang. Because of this very strong Big Bang-dark matter linkage, strong evidence of the existence of relativistic-proton dark matter would provide strong evidence for the existence of the Relativistic Big Bang. The author believes that his 2003 and 2006 books and his 2005 scientific paper provide very strong scientific evidence for the existence of relativistic-proton dark matter and, therefore, for the existence of the RBB.

Cosmological support for an RBB may come eventually via compatibility with, for example, the Cosmic Microwave Background, or Cosmic Inflation, or the Second Law of Thermodynamics, or the temperatures of the Big Bang, or the mass values for dark matter particles, or the 83% dark matter mass. Note that an RBB would be a very efficient way of creating the Universe and conserving its energy because the fewest number of particles and the least useless energy would be created and dispersed, which are characteristics that may be compatible with Cosmic Inflation theory and its associated Big Bang.

As previously indicated, strong scientific evidence that the dark matter of the Universe is comprised of galaxy-orbiting relativistic protons can be found in the 2003 book, "How Dark Matter Created Dark Energy and the Sun" *[5],* the 2005, 19-page scientific paper, "Identifying Dark Matter Through the Constraints Imposed by Fourteen Astronomically Based 'Cosmic Constituents'"*[9],* and the 2006 book, "Comprehending and Decoding the Cosmos: Discovering Solutions to Over a Dozen Cosmic Mysteries by Utilizing Dark Matter Relationism, Cosmology, and Astrophysics" *[8].*

Confirmation of the identification of dark matter is scientifically supported in the 2006 book through the utilization of the relativistic-proton dark matter hypothesis, in conjunction with the laws of physics, to derive solutions and plausible explanations for more than 15 previously unsolved cosmic mysteries.

If the existence of the relativistic-proton dark matter provides strong evidence that the Big Bang satisfied the Second Law of Thermodynamics, then a corollary could follow: Since the Big Bang must have satisfied the Second Law of Thermodynamics, its entropy must have been very low; and since relativistic protons possess the lowest possible entropy, they must have represented the principal mass output of the Big Bang.

**Acknowledgments**

I thank the British and Japanese for the weeks my books were on their Amazon.com's Best Seller lists. I also thank everyone involved in including my 2003 book among five books recommended for reading in cosmology in conjunction with Sir Martin Rees' TV science program, "What We Still Don't Know."

**References**

1. F. Zwicky, 1937 *Astrophys. J.* (Lett) 86, 217

2. V. C. Rubin, N. Thonnard and W. K.Ford, 1978 *Astrophys. J.* (Lett) 225 , L107

3. V. Rubin, *Bright Galaxies – Dark Matters* (Amer. Inst. Physics, New York, 1997), p. 109-116

4. G. Blumenthal, S. Faber, J. R. Primack, and M. J. Rees, 1984 *Nature* **311**, 517

5. J. Drexler, *How Dark Matter Created Dark Energy and the Sun* (Universal Publishers, Parkland, Florida, USA, 2003), p. 16

6. J. Drexler, *op. cit.*, p. 18

7. A. S. Bishop, *Project Sherwood – The U.S. Program in Controlled Fusion* (Addison- Wesley Publishing Company, Inc., Reading, Massachusetts, U.S.A. 1958) p 177-178

8. J. Drexler, *Comprehending and Decoding the Cosmos* (Universal Publishers, Boca Raton, Florida, USA, 2006)

9. J. Drexler, 2005, astro-ph/0504512 v1

10. S. Hawking, Lecture – Life in the Universe, http://www.hawking.org.uk/lectures/life.html

11. S. Koppes, The University of Chicago Chronicle, Vol. 24, No.5, November 18, 2004, http://chronicle.uchicago.edu/041118/entropy.shtml